\documentclass[aps,prl,twocolumn,groupedaddress,nofootinbib,showpacs]{revtex4-1}
\usepackage[dvips]{graphicx}
\newif\ifdraft
\drafttrue
\newif\ifpreprint
\preprinttrue

\def\fig#1{fig.~{\ref{#1}}}

\def\eqn#1{eq.~({\ref{#1}})}
\def\nn{\nonumber}
\def\NeqFour{\mathcal{N}=4}

\def\tree{{\rm tree}}
\def\dual{{\rm dual}}
\def\Tr{{\rm Tr}}
\def\ts{\textstyle}
\def\na{n}
\def\nb{n}

\newbox\charbox
\newbox\slabox
\def\s#1{{      
        \setbox\charbox=\hbox{$#1$}
        \setbox\slabox=\hbox{$/$}
        \dimen\charbox=\ht\slabox
        \advance\dimen\charbox by -\dp\slabox
        \advance\dimen\charbox by -\ht\charbox
        \advance\dimen\charbox by \dp\charbox
        \divide\dimen\charbox by 2
        \raise-\dimen\charbox\hbox to \wd\charbox{\hss/\hss}
        \llap{$#1$}
}}

\begin{document}

\title{
\ifpreprint
\hbox{\rm\small UCLA/11/TEP/102}
\fi
A Color Dual Form for Gauge-Theory Amplitudes
}
 
\author{Z.~Bern and T.~Dennen}

\affiliation{Department of Physics and Astronomy, UCLA, Los Angeles, CA
90095-1547, USA}

\begin{abstract}
Recently a duality between color and kinematics has been proposed,
exposing a new unexpected structure in gauge theory and gravity 
scattering amplitudes.  Here we propose that the relation goes
deeper, allowing us to reorganize amplitudes into a form
reminiscent of the standard color decomposition in terms of traces
over generators, but with the role of color and kinematics swapped.
By imposing additional conditions similar to Kleiss-Kuijf relations
between partial amplitudes, the relationship between the earlier form
satisfying the duality and the current one is invertible.  We comment
on extensions to loop level.
\end{abstract}

\pacs{04.65.+e, 11.15.Bt, 11.25.Db, 12.60.Jv \hspace{1cm}}

\maketitle


Gauge theories have been long studied as descriptions of fundamental
forces of Nature, offering detailed theoretical predictions for
experiments and observations.  In recent years, on the theoretical
side, a surprising array of simple structures have been uncovered in
gauge-theory scattering amplitudes, especially in maximally
supersymmetric gauge theories.  The first of these was
Witten's striking observation that amplitudes are supported on
curves in twistor space~\cite{WittenTopologicalString}. This was then
followed by an impressive list of newly uncovered symmetries and structures.

\begin{figure}[b]
\includegraphics[clip,scale=0.4]{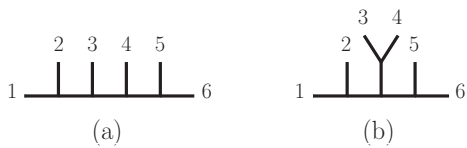}
\caption[a]{The two diagram types at six points.
Each graph can be taken to represent a color factor, a numerator or a set of
Feynman propagators. }
\label{SixPtFigure}
\end{figure}

The present Letter will focus on one of these recently uncovered structures, the BCJ duality between color and kinematics~\cite{BCJ,BCJLoop}. The duality is described in terms of graphs with only
cubic vertices.  In particular, at tree level we can decompose any
scattering amplitude of adjoint color representation states as
\begin{equation}
\mathcal{A}_m^{\rm YM} = g^{m-2}\sum_j\frac{c_j n_j}
{\prod_{\alpha_j} p_{\alpha_j}^2}\,,
\label{TreeDiagrams}
\end{equation}
where the sum runs over graphs, labeled by $j$.  At six points, for
example, there are 105 cubic graphs. The two basic ones are displayed
in \fig{SixPtFigure}; the others are given by 
relabelings of these.  The product in the denominator of
\eqn{TreeDiagrams} runs over the Feynman propagators corresponding to
each internal line of graph $j$.  The $c_j$ are the color factors
obtained by dressing every three vertex with an $\tilde f^{abc} = i
\sqrt{2} f^{abc}$ structure constant in the usual way, and the $n_i$
are kinematic numerator factors.  Representations of the form in
\eqn{TreeDiagrams} can be obtained from Feynman diagrams, or other
starting representations.  We assign terms to diagrams according to
the color factors.  If a term is missing a propagator of
the diagram to which it was assigned, we simply multiply and divide by
the appropriate factor of $p_{\alpha_i}^2$ to ensure all propagators
are present. The numerators do not necessarily have to be local, and
some can vanish; the key constraint is that a given choice of
numerators in \eqn{TreeDiagrams} yields the correct amplitude.

The BCJ duality proposes that there exist representations of the
amplitude such that for any set of three graphs $j_1,j_2,j_3$,
related by a color Jacobi identity, there is a corresponding numerator
relation,
\begin{equation}
 c_{j_1}\pm c_{j_2} \pm c_{j_3}=0\; \Rightarrow \;n_{j_1}\pm n_{j_2}\pm n_{j_3}=0\,,
\label{Jacobi}
\end{equation}
where the relative signs are dictated by the choice of signs in
defining the color factors.  In addition, the $n_j$ are required to
satisfy the same antisymmetry relations as satisfied by color factors
under any relabelings,
\begin{equation}
c_j \rightarrow - c_j \; \Rightarrow \; n_j \rightarrow -n_j\,.
\label{AntiSymmetry}
\end{equation}
The $n_i$ do not necessarily need to be manifestly crossing symmetric.
Indeed, solutions of the $n_j$ in terms of amplitudes, which
violate this condition, may be found in
refs.~\cite{BCJ,Michael}.  The conjecture has also been extended
to all loop orders~\cite{BCJLoop}, offering  a rather simple
direct construction of gravity loop amplitudes from corresponding
gauge-theory ones.
One consequence of this duality is that it implies nontrivial
relations between the tree-level color-ordered partial amplitudes of
gauge theory~\cite{BCJ}.  These properties have also been studied from
the vantage points of string theory~\cite{BCJAmplStrings} and
field theory~\cite{BCJAmplRelations, Square}. 

How far does the analogy between color and kinematics extend?  We know
that in SU$(N_c)$ gauge theory useful trace representations of the
color factors exist.  In the present Letter we show that the analogy
between color and kinematics is sufficiently robust that a
representation of the kinematic numerators exists which shares the
same algebraic properties as color traces.  Moreover we will show that
additional interesting constraints can be imposed which uniquely
determine the kinematic trace-like representation in terms of 
kinematic numerators satisfying the duality.

At tree level, the well-known trace-based color decomposition of gluon
amplitudes is~\cite{TreeReview},
\begin{equation}
\hskip -.01 cm 
{\cal A}_m^{\rm tree}  = 
g^{m-2} \sum_{\sigma} \Tr[T^{a_1} \cdots T^{a_m}] 
A_m^\tree (1,2, \ldots, m)\,,
\label{FundColorRep}
\end{equation}
where $g$ is the gauge coupling, $A_m^\tree$ is a partial amplitude
stripped of color factors, and the $T^{a_i}$ are
fundamental-representation matrices of an SU$(N_c)$ Lie group.  The
sum runs over all non-cyclic permutations of external legs.  The
labels on momenta, polarizations or spinors, implicit in
\eqn{FundColorRep}, are also to be permuted in the sum.  The
color-stripped partial amplitudes can be expressed as a subset of
diagrams following the same ordering of legs as in the partial
amplitudes, but with no color factors.  (For example, see eq.~(4.5) of
ref.~\cite{BCJ}.)

\begin{figure}[tb]
\includegraphics[clip,scale=0.4]{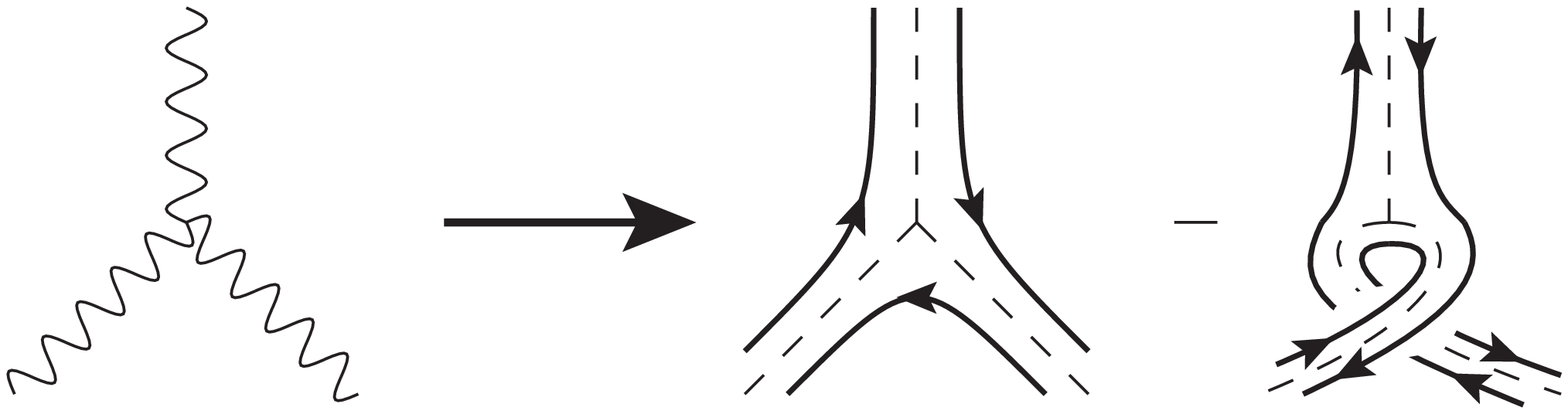}
\caption[a]{An antisymmetric vertex in a cubic graph is replaced by a
  difference of two double-line vertices.}
\label{ToPThreeFigure}
\end{figure}

We propose that a dual description exists where we can
swap the role of color and kinematics in the trace-based 
color decomposition, in particular by
rewriting \eqn{FundColorRep} in a dual form,
\begin{eqnarray}
{\cal A}_m^{\rm tree} &=& 
g^{m-2} \sum_{\sigma}\tau_{(12\ldots m)}
A_m^\dual (1,2, \ldots, m)\,,
\label{DualFundRep}
\end{eqnarray}
where the $\tau_{(12\ldots m)}$ are kinematic prefactors
satisfying the same cyclic properties as color
traces.   $A_m^\dual$ is a dual amplitude
defined by replacing all kinematic
numerators with color factors.  That is, they can be generated by the
same color-ordered graphs that generate $A_m^{\rm tree}$, except instead of
a kinematic factor at every vertex, we simply have an $\tilde f^{abc}$.  
The dual amplitudes are not gauge-theory amplitudes, but amplitudes
in a $\phi^3$ theory dressed with group-theory factors.

If we assume that the duality~(\ref{Jacobi}) holds, then gravity
amplitudes can be obtained directly from Yang-Mills numerators by
replacing the color factors with another copy of the kinematic
numerators~\cite{BCJ}, as proven at tree level~\cite{Square} and
conjectured to hold to all loop orders~\cite{BCJLoop}.  Since the
kinematic numerators share the same algebraic properties as color
factors, it is then straightforward to rearrange gravity amplitudes
into a form analogous to the gauge-theory dual form
(\ref{DualFundRep}),
\begin{eqnarray}
{\cal M}_m^{\rm tree} &=& 
i \biggl(\frac{\kappa}{2}\biggr)^{m-2
} \sum_{\rm \sigma}\tau_{(12 \ldots m)} 
 A_m^\tree (1,2, \ldots, m)\,,\nn
\label{GravFundRep}
\end{eqnarray}
Here $\kappa$ is the gravitational coupling, $ A_m^\tree$ are partial
amplitudes of Yang-Mills theory and $\tau$ is exactly the same
kinematic prefactor as in \eqn{DualFundRep}.

The $\tau$'s are generated by expressing each numerator 
in terms of a set of objects which satisfy the cyclic
symmetry of color traces. For example, at the three-point level
we demand that, 
\begin{eqnarray}
n_{123}=\tau_{(123)}-\tau_{(132)} \,, \nn
\end{eqnarray}
where $n_{123}$ is just the three vertex, as illustrated in
\fig{ToPThreeFigure}.  In general, we will use parentheses around the
subscript labels on the $\tau$'s to indicate which color trace the
quantity is analogous to.  In particular, $\tau_{(123)}$ is analogous
to $\Tr[T^{a_1} T^{a_2} T^{a_3}]$.

\begin{figure}[tb]
\includegraphics[clip,scale=0.35]{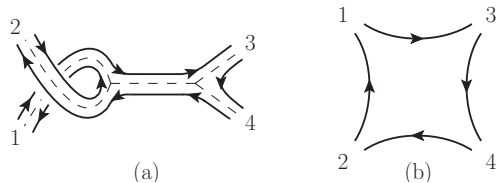}
\caption[a]{Sewing of two vertices in a double-line graph (a).  The
  ordering of the external legs follows the arrow
  around the graph. This graph corresponds with the kinematic quantity
  $\tau_{(1342)}$. The same double-line graph is displayed in (b) in a
  form emphasizing that it is the same quantity whether we sew the two
  three-point $\tau$'s in the $12$ channel or  $13$ channel.
}
\label{DoubleLineSewingFigure}
\end{figure}

For any number of legs, we can associate diagrams to the $\tau$'s in a
manner completely parallel to the standard 't~Hooft double-line
formalism for color. Just as the color traces ``trivialize'' the color
Jacobi identities, the $\tau$'s will do the same for the kinematic
numerator factors, emphasizing the parallelism between color and
kinematics.  For example, in \fig{DoubleLineSewingFigure}(a) we
display the double-line diagram for $\tau_{(1324)}$ obtained by sewing
together two double-line three-point graphs.  For the duality
(\ref{Jacobi}) to hold, the kinematic expression associated with each
double-line graph should depend on only the topological structure of
the graph, rather than on the detailed structure of vertices and
internal lines in the underlying cubic graph.  That is, a more
appropriate way to draw the graph in \fig{DoubleLineSewingFigure}(a)
is shown in \fig{DoubleLineSewingFigure}(b). In much the same way as
single traces used in the tree-level color decomposition depend on
only the cyclic ordering of legs, we demand that the $\tau$'s also
depend on only the ordering.  The property that one should obtain the
same object when sewing in either channel is of course reminiscent of
a key feature of string theory.  However, at present we take the
diagrams only as guides, since we do not have rules for directly
combining lower-point $\tau$'s into higher-point ones.

To be more explicit, consider some examples.
At four points there are three graphs contributing to \eqn{TreeDiagrams}.
These numerators are expressed in terms of $\tau$ via
\begin{equation}
\hskip -.1 cm 
 n_{12(34)} = \tau_{(1[2,[3,4]])}\,,
\label{4ptdecomp}
\end{equation}
where the parenthesis on the indices of $n$ indicates the associated
propagators, {\it i.e.} in this case we have one, $i/(k_3 + k_4)^2$.
For $\tau$ the brackets signify an antisymmetric combination, {\it
  i.e. } $\tau_{(12[3,4])} \equiv \tau_{(1234)}-\tau_{(1243)}$.  The
two other channels at four points are just relabelings of this
channel.  We note that by expressing the $n_j$ in terms of $\tau$, the
Jacobi-like equation $n_{12(34)}-n_{23(41)}-n_{42(31)} =0$ holds
automatically.

Can we find an explicit form of $\tau$ with the desired
cyclicity?  It is not difficult to check at four points that
\begin{eqnarray}
 \tau_{(1234)} &=& \frac{1}{6}(n_{12(34)} + n_{23(41)})  \hskip .5 cm  \nn
\end{eqnarray}
indeed satisfies cyclicity and after using the duality relations
(\ref{Jacobi}) returns the numerators when combined as in \eqn{4ptdecomp}. The other $\tau$'s are given by
relabelings.  Interestingly, this solution satisfies some additional
properties, namely invariance under a reversal of arguments 
and also an identity reminiscent of
the U$(1)$ decoupling identity for amplitudes,
\begin{eqnarray}
 \tau_{(1234)} + \tau_{(1342)} + \tau_{(1423)} = 0\,. \nn
\end{eqnarray}

More generally, our ability to express the
$\tau$'s directly in terms of the graph numerators $n_j$ is precisely
dependent on having $\tau$ satisfying identities of the same form as
Kleiss-Kuijf identities,
\begin{eqnarray}
\tau_{(1\{\alpha\}m\{\beta\})} &=& 
(-1)^{|\beta|} \sum_{\{\sigma\}} \tau_{(1\{\sigma\}m)},
\label{KleissKuijfLike}
\end{eqnarray}
where the sum is over the ``ordered permutations'' $\{\sigma\} \in
{\rm OP}(\{\alpha\}, \{\beta^T\})$, that is, all permutations of
$\{\alpha\} \bigcup \, \{\beta^T\}$ that maintain the order of the
individual elements belonging to each set within the joint set.  The
notation $\{\beta^T\}$ represents the set $\{\beta\}$ with the
ordering reversed, and $|\beta|$ is the number of elements in
$\{\beta\}$.  For gauge-theory partial amplitudes, these relations
were conjectured in ref.~\cite{KleissKuijf} and proven in
ref.~\cite{LanceColor}.  They are a consequence of the antisymmetric
nature of the vertices describing the $n_j$, as noted in
ref.~\cite{BCJ}.  Indeed, any cyclic object, such as $\tau$, that can
be expressed as a linear combination of the $n_j$ with prefactors that
respect the symmetry and relabeling properties of the $n_j$
automatically will satisfy the Kleiss-Kuijf relations.

At five points we have
\begin{eqnarray}
 n_{12(3(45))} &=& \tau_{(1[2,[3,[4,5]]])} \,,\nn\\
 \tau_{(12345)} &=& \frac{1}{20} \sum_{\sigma} n_{12(3(45))} \,, \nn
\end{eqnarray}
where the sum runs over cyclic permutations.  The numerator
$n_{12(3(45))}$ is the graph with Feynman propagators $i/(k_1+k_2)^2$
and $i/(k_4+k_5)^2$.  One can straightforwardly verify that
this $\tau$ satisfies the relations
(\ref{KleissKuijfLike}).

At six points we can express the numerators of the two diagrams shown 
in \fig{SixPtFigure}(a) and (b) in terms of the $\tau$ via
\begin{eqnarray}
\na_{12(3(4(56)))} &\!=\!& \tau_{(1[2,[3,[4,[5,6]]]])} \,,\nn\\
\nb_{(12)(34)(56)} &\!=\!& \na_{12(3(4(56)))} - \na_{12(4(3(56)))}\,. \nn
\end{eqnarray}
The decomposition of $\tau$ in terms
of the numerators is more complicated, in part because non-trivial
rearrangements are possible using the duality  (\ref{Jacobi}).  
One such solution is
\begin{eqnarray}
\tau_{(1\ldots6)}\nn
&\!=\!& {\ts \frac{1}{1890}} \sum_\sigma \Bigl(
 32 \na_{12(3(4(56)))} - 3 \na_{12(4(3(56)))} \nn \\
&& \null \hskip -0.6 cm 
 -{\ts \frac{31}{2}} \na_{12(3(6(45)))}
 -{\ts \frac{31}{2}} \na_{12(6(3(45)))}
 + 2 \na_{36(1(2(45)))}\nn \\
&& \null \hskip -0.6 cm 
 + 2 \na_{36(2(1(45)))} 
 + 2 \na_{36(4(1(25)))} - \na_{26(1(4(35)))}\nn\\
&& \null \hskip -0.6 cm 
 - \na_{26(4(1(35)))} 
 - \na_{35(1(2(46)))} - \na_{35(2(1(46)))}\nn\\
&& \null \hskip -0.6 cm  
 + \na_{24(1(3(56)))} + \na_{24(3(1(56)))}
 - \na_{26(1(3(45)))}\nn\\
&& \null \hskip -0.6 cm 
 - \na_{26(3(1(45)))} \Bigr)\,,  
\label{tau6Def}
\end{eqnarray}
where here the sum runs over the cyclic permutations of labels. 
The reader may also
verify that $\tau_{(1\ldots6)}$ satisfies the Kleiss-Kuijf
relations~(\ref{KleissKuijfLike}), given the algebraic properties of
the kinematic numerators.

We have explicitly verified through nine points that an expression
for $\tau$ in terms of kinematic numerators (as in \eqn{tau6Def})
exists and that it automatically satisfies the Kleiss-Kuijf-like
relations (\ref{KleissKuijfLike}).  The explicit expression for $\tau$
at $m=7$ contains more than $600$ numerators and grows rapidly 
as the number of legs increases.  Generally, it is better 
to think of the $n_j$ numerators as functions of the $\tau$.  A general
solution is
\begin{eqnarray}
\nonumber
n_{12(3( \ldots m)) \cdots)} = \tau_{(1[2,[3,[ \ldots m]] \cdots])} \,.
\end{eqnarray} 
The remaining numerators can be obtained by solving the duality
relations (\ref{Jacobi}), because they automatically hold for
numerators expressed in terms of the $\tau$'s.

An interesting consequence of the above is that we can define an
alternative color decomposition in place of \eqn{FundColorRep}, where
instead of color traces we use objects that satisfy Kleiss-Kuijf
relations as well.  These objects are given by taking the above
equations for the $\tau$ in terms of the numerators $n_j$ and
replacing them with color factors $c_j$.  This gives a valid set of
objects to use in place of color traces in \eqn{FundColorRep}.
Using the explicit formulas given
above, it is straightforward to confirm though six points that the
amplitude in \eqn{FundColorRep} is unchanged under this substitution.


Can this construction be extended to loop level?  As an initial
peek at this question, we turn to the simple case of a one-loop
four-point $\NeqFour$ super-Yang-Mills amplitude, first obtained in
ref.~\cite{GSB}.  We follow the same diagrammatic double-line
formalism as at tree level.

\begin{figure}[tb]
\includegraphics[clip,scale=0.4]{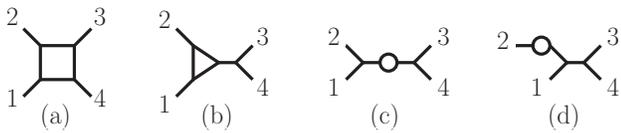}
\caption[a]{Cubic diagrams appearing in one-loop four-point amplitudes.}
\label{1loopgraphs}
\end{figure}

The four types of cubic diagrams contributing to the four-point one-loop
amplitude are shown in \fig{1loopgraphs}. The numerators associated
with these diagrams in the BCJ representation are
\begin{eqnarray}
  n_{(\rm{a})} = stA^{\tree}\,,\qquad
  n_{(\rm{b})} = n_{(\rm{c})}=n_{(\rm{d})}=0 \,, \nn
\end{eqnarray}
where $s$ and $t$ are standard four-point Mandelstam invariants.
Although this seems like a trivial state of affairs, expanding each
numerator in terms of the double-line graphs reveals unexpected
structure. In this expansion, there are multiple lines flowing
around the double-line graphs, and we indicate the external legs
attached to each line with a $(12\ldots)$ in the subscript of
$\tau$. Since each one-loop graph carries two independent lines,
each $\tau$ will have two sets of parentheses in the subscripts.
The graphs that appear in the expansion of the present example
are $\tau_{(1)(234)}$, $\tau_{(12)(34)}$ and $\tau_{(1234)()}$, along
with relabelings of these. The decomposition of the four numerators in
terms of the $\tau$'s is straightforward and closely parallels
a U$(N_c)$ color decomposition.

One immediate solution to the decomposition can be obtained by setting
$\tau_{(1234)()}$ proportional to the box numerator, and the other two
$\tau$ functions to zero. A similar construction also works for the
five-point amplitude~\cite{UnitarityMethod} of $\NeqFour$ super-Yang-Mills
theory.  In this case $\tau_{(12345)()}$ are set
proportional to the pentagon numerator given in
ref.~\cite{Cachazo}.\footnote{We thank J.~J.~M.~Carrasco and
H.~Johansson for pointing out that the color-kinematics duality holds
with this form.}

A more interesting solution to the four-point decomposition is
\begin{eqnarray}
 \tau_{(1234)()} &=& \frac{1}{62}stA^{\tree}\,, \hskip .6cm 
 \tau_{(12)(34)} = \frac{3}{31}stA^{\tree}\,, \nn\\
 \tau_{(1)(234)} &=& -\frac{3}{62}stA^{\tree}\,. \nn
\end{eqnarray}
With this solution the $\tau$ functions satisfy the same identities 
as the color-ordered partial amplitudes, namely,
\begin{eqnarray}
  \tau_{(\{\alpha\})(\{\beta\})} &=& (-1)^{|\beta|}
\sum_{\{\sigma\}} \tau_{(\{\sigma\})()}\,, \nn
\end{eqnarray}
where the sum is over the ``cyclically ordered permutations'' ${\rm COP}(\{\alpha\},\{\beta^T\})$, that is, all permutations
of $\{\alpha\}\bigcup\,\{\beta^T\}$ that maintain the cyclic
orderings of $\{\alpha\}$ and $\{\beta^T\}$ separately, and with one
leg fixed (see ref.~\cite{UnitarityMethod}).

To go beyond these simple examples, one needs a self-consistent
assignment of loop momentum labels, with an understanding of the
proper way to associate kinematic information with the double-line
graphs.  At tree and one-loop level, each double-line graph can
be interpreted as defining an embedding of the underlying cubic graph
in a plane, while at higher loops, non-planar embeddings also appear,
introducing a topological hierarchy to the graphs analogous to the
$1/N_c$ expansion of Yang-Mills.  We leave the study of loop level to
future work.


There are also a number of other interesting open questions.  For
example, it would be very useful to find a direct recursive procedure
for building the $\tau$'s.  Such a construction would automatically
produce numerators that satisfy the color-kinematics duality.  We also
note that the double-line graphs describing the kinematic trace
representation are reminiscent of open string diagrams.  This brings
up the interesting question of whether the properties we have
described here can be unraveled in string theory.  More generally, the
trace-like representation described here emphasizes a group-theoretic
origin for the duality.  Because the same kinematic numerators appear
in gravity theories, the same underlying group theory should carry
over to gravity.  It remains an important challenge to understand the
origin of this duality and to fully map out its implications.

\vskip .3 cm 

We thank N.~Arkani-Hamed, J.~J.~M.~Carrasco, Y.~t.~Huang, H.~Ita,
H.~Johansson and K.~Ozeren for helpful discussions.  This research was
supported by the US Department of Energy under contract
DE--FG03--91ER40662.

\end{document}